\documentclass[preprint,12pt]{elsarticle}

\usepackage{amssymb}

\biboptions{sort&compress}

\journal{Nuclear Instruments and Methods in Physics Research A}

\begin{document}

\begin{frontmatter}



\title{Neutron Beam Tests of Barium Fluoride Crystal for Dark Matter Direct Detection}


\renewcommand{\thefootnote}{\fnsymbol{footnote}}
\author{
C.~Guo$^{a,b}$,
X.H.~Ma$^{a,}$\footnote{Corresponding author: Tel:~+86-1088233167. E-mail address: maxh@ihep.ac.cn (X.H.~Ma).},
Z.M.~Wang$^{a}$,
J.~Bao$^{c}$,
C.J.~Dai$^{a}$,
M.Y.~Guan$^{a}$,
J.C.~Liu$^{a}$,
Z.H.~Li$^{a}$,
J.~Ren$^{c}$,
X.C.~Ruan$^{c}$,
C.G.~Yang$^{a}$,
Z.Y.~Yu$^{a}$,
W.L.~Zhong$^{a}$
}

\address{
${^a}$Key Laboratory of Particle Astrophysics, Institute of High Energy Physics, Chinese Academy of Science,Beijing, China\\
${^b}$ School of Physics, University of Chinese Academy of Science, Beijing, China \\
${^c}$ Science and Technology on Nuclear Data Laboratory, China Institute of Atomic Energy, Beijing, China\\
}

\begin{abstract}
In order to test the capabilities of Barium Fluoride (BaF$_{2}$) Crystal for dark matter direct detection, nuclear recoils are studied with mono-energetic neutron beam. The energy spectra of nuclear recoils, quenching factors for elastic scattering neutrons and discrimination capability between neutron inelastic scattering events and ${\gamma}$ events are obtained for various recoil energies of the F content in BaF$_{2}$.
\end{abstract}


\end{frontmatter}


\section{Introduction}

Barium fluoride (BaF$_{2}$) scintillator detectors are in common use for the measurement of $\gamma$s over a wide energy range, e.g., in the time-of-flight method for the localization of positron-emitting radionuclides in nuclear medicine, because of the relatively high detection efficiency and the fast response which allows subnanosecond timing ~\cite{1983-NIMA-timing,1984-NIMA-timing,1987-NIMA-timing}. The BaF$_{2}$ scintillation light primarily consists of fast and slow components of which the light emitting time constants are 0.9~ns and 650~ns~\cite{2014-PDG} respectively, and the intensity ratio of each component is sensitive to the incident radiation type: the fast component is larger for $\gamma$s, but much smaller for charged particles including alpha, proton, deuteron, etc. ~\cite{1986-NIMA-242-charged-gamma,1986-NIMA-253-charged-gamma,1987-NIMA-254-charged-gamma,1987-NIMA-262-charged-gamma,2002-NIMA-alpha-gamma}. This feature of BaF$_{2}$ makes it possible to discriminate charged particles from $\gamma$s by using pulse shape discrimination (PSD) techniques. Furthermore, when reacting in BaF$_{2}$, high energy neutrons (10~MeV-150~MeV) generate secondary light charged particles so that they can be discriminated from $\gamma$ interactions by using PSD techniques~\cite{1989-NIMA-274-neutron,1989-NIMA-285-neutron,1994-NIMA-346-neutron}. It makes BaF$_{2}$ widely used in the measurement of neutron capture cross sections which is the most important input data of nuclear astrophysics~\cite{1990-PRC-4pi}.

On the other hand, there was not any attempt to measure nuclear recoil of BaF$_{2}$ from neutrons. The measurement becomes essential at present because BaF$_{2}$ is realized to be an important sensitive material for dark matter direct detection in an underground laboratory. In BaF$_{2}$, $^{19}F$ has the highest spin factor and then the largest cross section so that it can provide the highest event rate in spin-dependent (SD) dark matter elastic scattering~\cite{SD1,SD2,SD3,SD4,SD5}. Several experiments have focused on~\cite{F19piccaso,F19simple,F19elegent,F19DAMA_1,F19DAMA_2} or will aim at~\cite{F19simpleIII,F19mimac,F19dmtpc} the target material containing $^{19}F$. Besides, it is possible to detect a dark matter candidate through the inelastic scattering processes on the electron or on the nucleus targets~\cite{inel1,inel2,inel3}. BaF$_{2}$ has capability of making such detection, while the residual contamination of BaF$_2$ is the real issue that prevented it from being used in low background measurements~\cite{coin_1,coin_2}. We report our neutron beam test of BaF$_{2}$ in this paper.

\section{Experimental Set-up}

The test of nuclear recoils of BaF$_{2}$ is performed by using a neutron generator at the China Institute of Atomic Energy(CIAE). A pulsed mono-energetic 14.7~MeV neutron beam is generated from  T(D,n)${\alpha}$ reaction, which was also used to test nuclear recoil of CsI(Na) and CaF$_{2}$(Eu) in \cite{ourbeam}. The experimental set-up including the detector configuration, the electronics system and the analysis method was described in detail in \cite{ourbeam}. The sample of BaF$_{2}$, with all surface polished, is 2.5 $\times$ 2.5 $\times$ 2.5 $cm^3$ in cubic shape and is produced by the Beijing Glass Research Institute. The photomultipliers(PMT), 9821QB~\cite{ET} from Electron Tubes company(ET), has a quartz window so as to have very low radioactivity background and high quantum efficiency for BaF$_2$ emission light~\cite{emission}. Two PMTs directly face the top and bottom surfaces of the crystal, while the other four surfaces are wrapped by a 65~$\mu$m thickness Enhanced Specular Reflector(ESR) film. In order to identify neutrons scattered from the crystal, three neutron detectors (ND) are positioned at various angles and 1~m  away from the crystal sample (Table~\ref{table1}). Each neutron detector is made of liquid scintillator (BC501A) and one XP2020 PMT. The signals from the two crystal PMTs are sent to Flash Analog Digital Converter (FADC, CAEN V1729A, 1.25~$\mu$s readout window) to record the pulse shape. In the ND electronics system, the constant fraction timer(CFT) and the pulse shape discriminator(PSD CANBERRA 2160A) are used to discriminate neutrons from ${\gamma}$s by outputting square pulses, of which ${\gamma}$s have a smaller amplitude while neutrons have a larger one. The logic AND of the two crystal PMTs and the logic OR of all the NDs are performed, and then the coincidence between them provides the trigger of the experiment. The beam start time is directly connected to the FADC for pulses record.

\begin{figure*}[htb]
  \centering
  \includegraphics[width=3in]{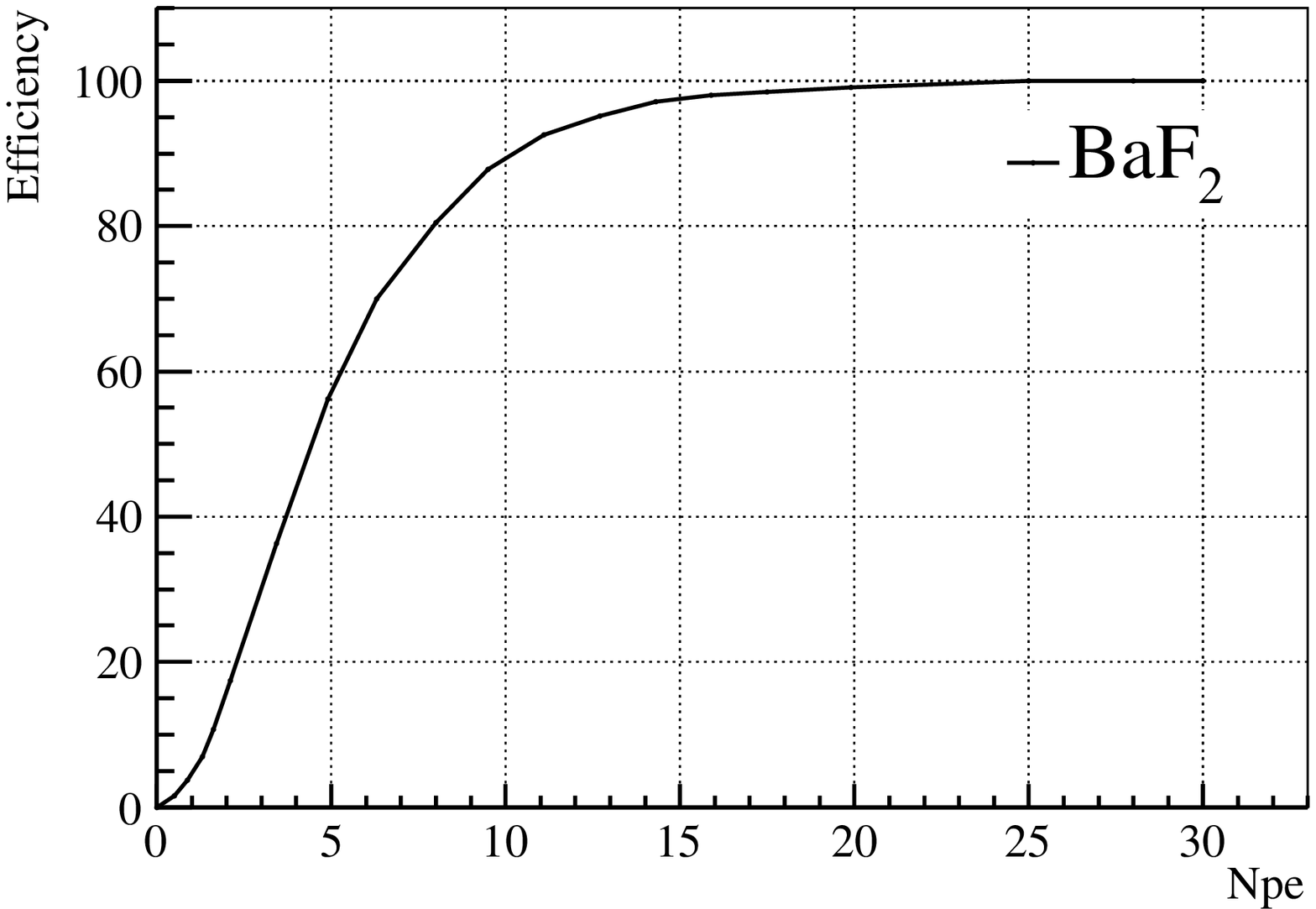}
     \caption{The trigger efficiency of BaF$_2$ at different energies.}
  \label{BaF2_trigger}
 \end{figure*}

In order to reduce the ${\gamma}$ background induced by the neutron beam, the width of gate from the pulses are set to 100~ns and the width of the final trigger is also set to 100~ns. For the crystal detector, the trigger is formed by the coincidence of the two PMTs of which the threshold for each channel is 0.5 photoelectrons(p.e.). The trigger efficiency is mainly related to the crystal light emitting time constant and coincidence window. The trigger rate is $\sim$1~Hz while the expected event rate is $<$0.1~Hz. A Toy Monte Carlo is constructed to simulate the trigger efficiency, including the effects of single channel threshold, different crystal light emitting time constant and different coincidence window length. The result is also crosschecked with the measured spectra of $\gamma$ sources and background. Fig.~\ref{BaF2_trigger} shows the calculated trigger efficiency at different energies for BaF$_2$, where the trigger threshold is around 5~p.e. with $\sim$50\% efficiency at around 10~p.e.. To reject afterglows and Cherenkov pulses in the crystal, a 10~${\mu}$s veto is applied after each trigger. Table~\ref{table1} details the scattering angles and corresponding calculated recoil energies of which the recoil energy uncertainties are propagated from the NDs' position uncertainties and neutron energy uncertainties.

\begin{table}[h]
\begin{center}
\begin{tabular}{|l|c|c|c|}
\hline                      & ND1 & ND2 & ND3 \\ \hline
Scattering angle ($^{\circ}$)   &   30 ${\pm1}$  &   40 ${\pm1}$ &   25 ${\pm1}$  \\ \hline
recoil energy of Ba (keV)       & $29.4^{+1.9}_{-1.8}$  & $51.6^{+2.5}_{-2.4}$ & $20.5^{+1.6}_{-1.4}$  \\ \hline
recoil energy of F  (keV)       & $212.3^{+14.1}_{-13.7}$  &  $369.6^{+18.1}_{-16.9}$ &  $148^{+12.0}_{-11.4}$  \\ \hline
\end{tabular}
\caption{Scattering angles and estimated recoil energies for the three NDs.}
\label{table1}
\end{center}
\end{table}

\section{Analysis}

\begin{figure*}[!t]
  \centering
  \includegraphics[height=3in,angle=-90]{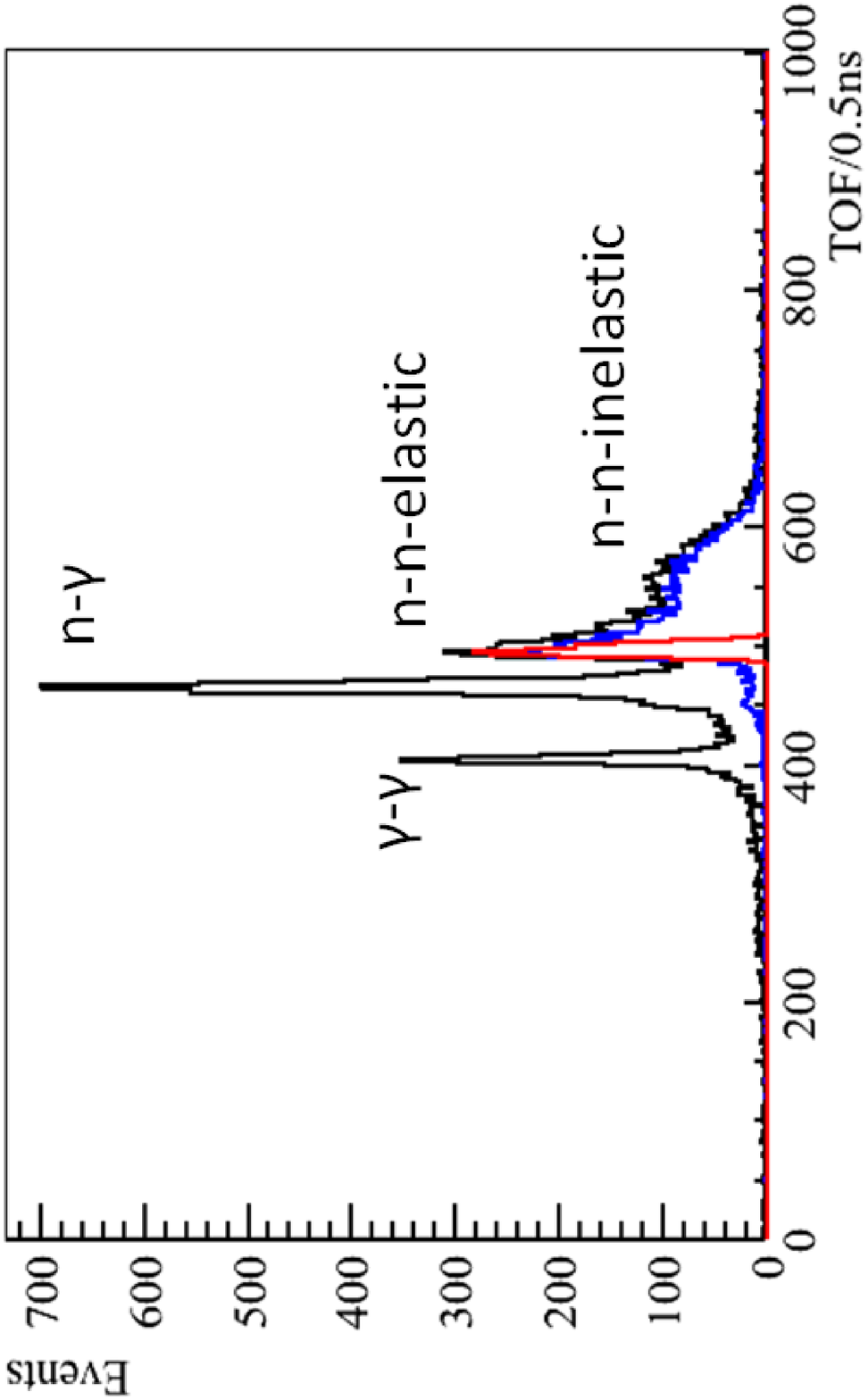}
  \includegraphics[height=3in,angle=-90]{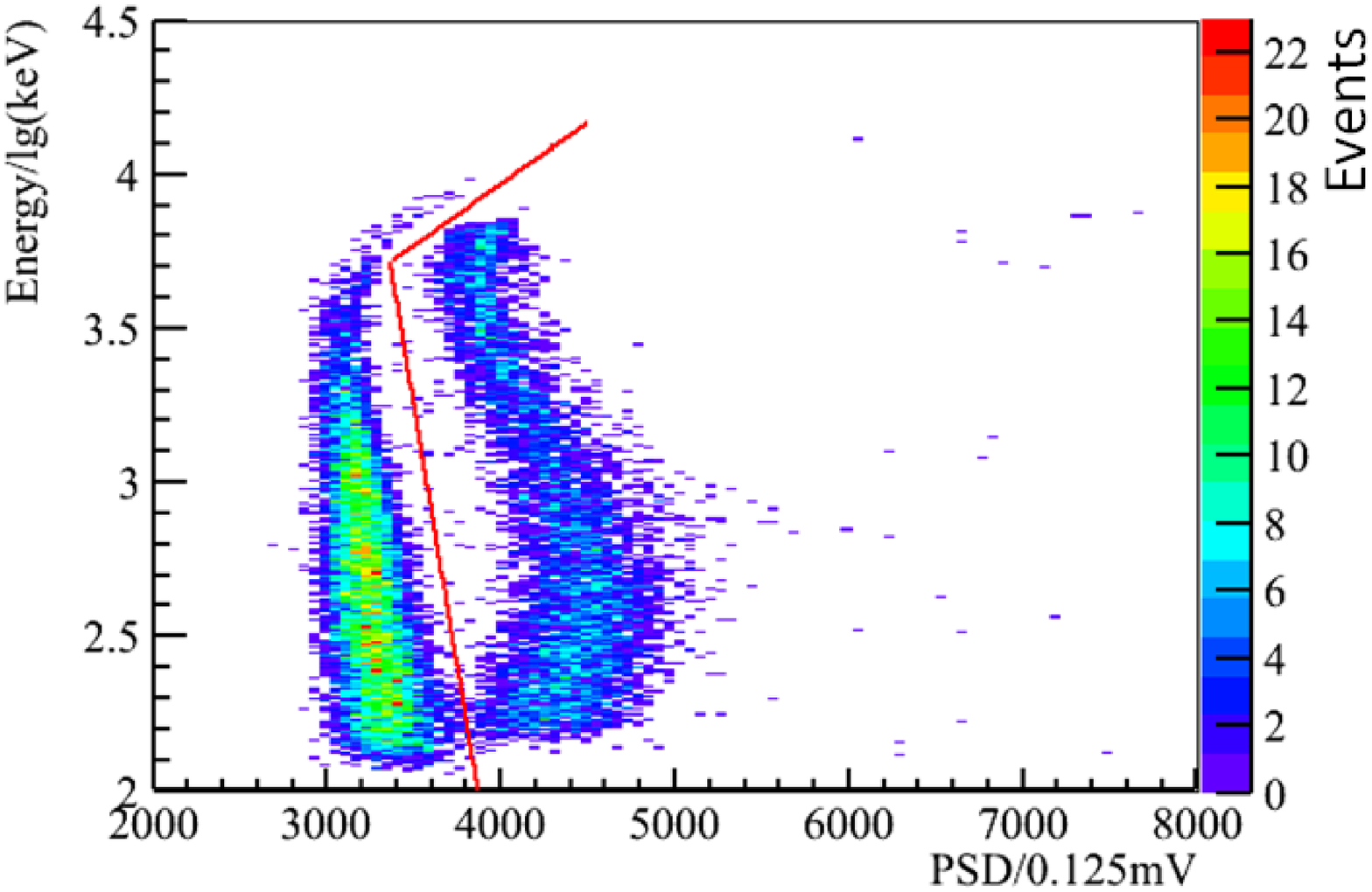}
  \includegraphics[width=3in]{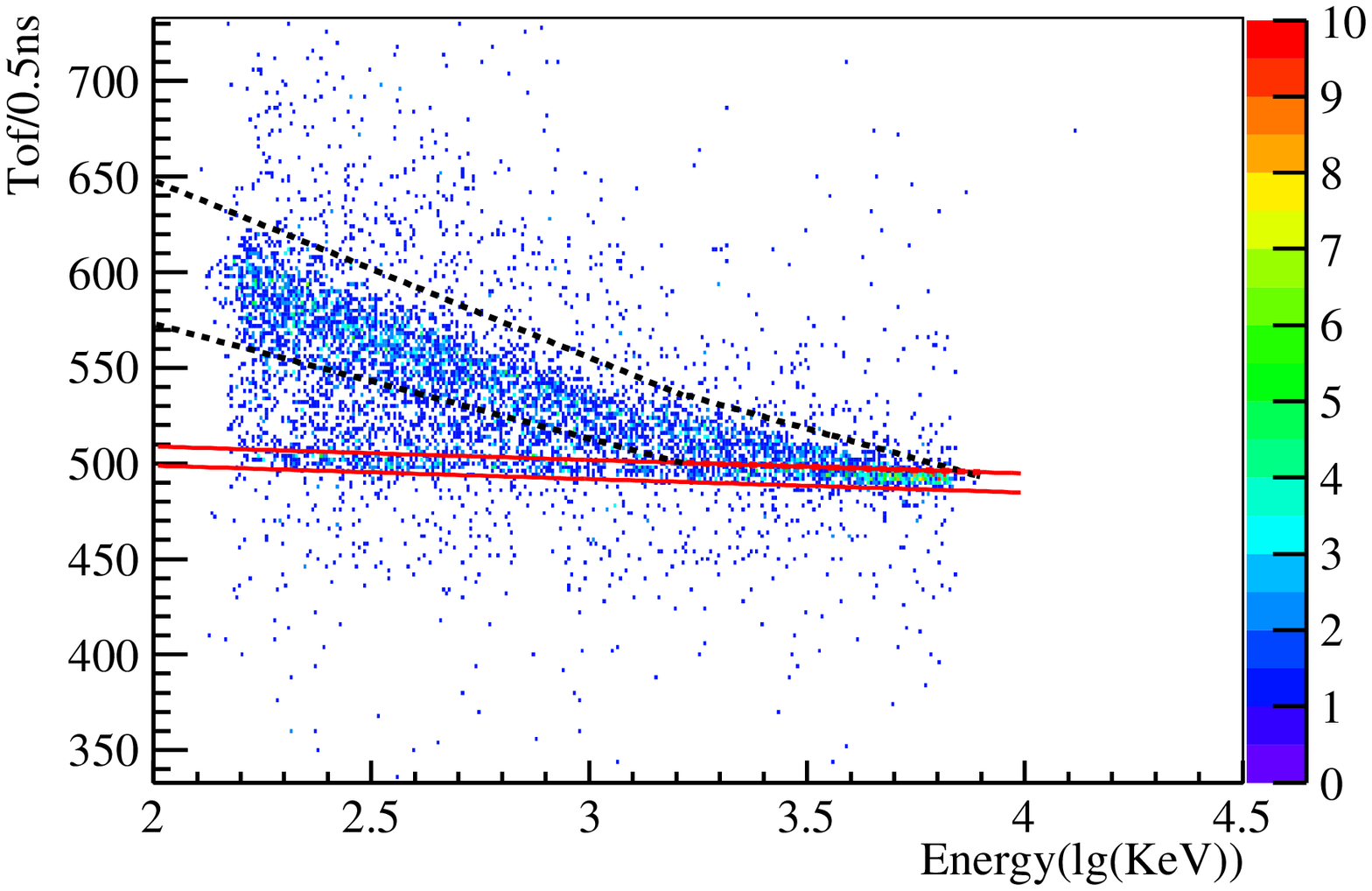}
    \caption{Upper: TOF distribution of BaF$_2$ from neutron source to ND1. Black line: After the Npe ratio cut selection; Blue line: After ND deposited energy VS PSD selection(the middle plot); Red line:After TOF VS ND deposited energy selection(the lower plot).  Middle: ND1 deposited energy VS PSD signal height. The red line is the separation between neutron events(right) and $\gamma$ events(left). Lower: TOF VS ND deposited energy. Region A: Elastic scattering events(between the two red lines); Region B: Inelastic scattering events(between the two black lines); Region C: Random coincident events. }
  \label{TOF}
 \end{figure*}

 The fluctuation of the measured time of flight(TOF) from neutron beam start time to NDs is 2.5~ns. TOF distribution of BaF$_2$ triggered with ND1(Fig.~\ref{TOF}, upper plot, black line) clearly has four peaks from left to right:

 \begin{enumerate}
 \item The ${\gamma}$-${\gamma}$ peak: The neutron beam is accompanied with amount of ${\gamma}$s which trigger ND1 after scattering with the crystal. Because $\gamma$ has the highest and fixed speed, this peak is on the first left and has the narrowest width. As different electronics and different length of cables the signals of NDs and beam go through, the zero of TOF is shifted to around 200ns.
 \item The n-${\gamma}$ peak: Part of neutrons react with BaF$_2$ crystal via inelastic processes and the generated ${\gamma}$s trigger ND1.
 \item The n-n-elastic peak: Part of neutrons are elastic scattered by the nuclei of BaF$_2$ crystal and the neutrons finally reach ND1. Because elastic scattering has a mono energy at the fixed scattering angle, this peak has a narrow width.
 \item The n-n-inelastic peak: Part of neutrons are inelastic scattered by the nuclei of BaF$_2$ crystal and the neutrons finally reach ND1. Because the energy losses of the inelastic scattered neutrons are higher than the elastic scattered neutrons and not mono-energetic, this peak is on the first right and has the widest distribution.
 \end{enumerate}

In order to select the scattering neutron events, the pulse shape discrimination selection by using neutron energy deposited in ND1 and PSD is performed (the middle plot of Fig.~\ref{TOF} ). After it, the scattering neutron events are clearly selected (Fig.~\ref{TOF}, upper plot, the blue line).

In order to discriminate elastic scattering and inelastic scattering neutron events, the selection by using neutron energy deposited in ND1 and TOF is performed. In the lower plot of Fig.~\ref{TOF}, the neutron events can be classified into three regions:
\renewcommand{\labelenumi}{\Alph{enumi}.}
 \begin{enumerate}
 \item	During elastic scattering with the crystal, neutrons lose a few energy in some keV(compared with the original energy of 14.7~MeV), so the speed of the neutrons is almost fixed except for a slight decrease along with the neutron energy deposited in ND1. Consequently the elastic scattering neutrons fall in the narrow area between the two parallel red lines.
 \item	During the inelastic scattering process, neutrons lose much energy in a wide range, so the speed of the neutrons and the TOF decrease along with neutron energy deposited in ND1, and they stay in area B above area A.
 \item	The randomly coincident background events uniformly spread in the whole area.
 \end{enumerate}

\begin{figure*}[!t]
  \centering
  \includegraphics[width=3in]{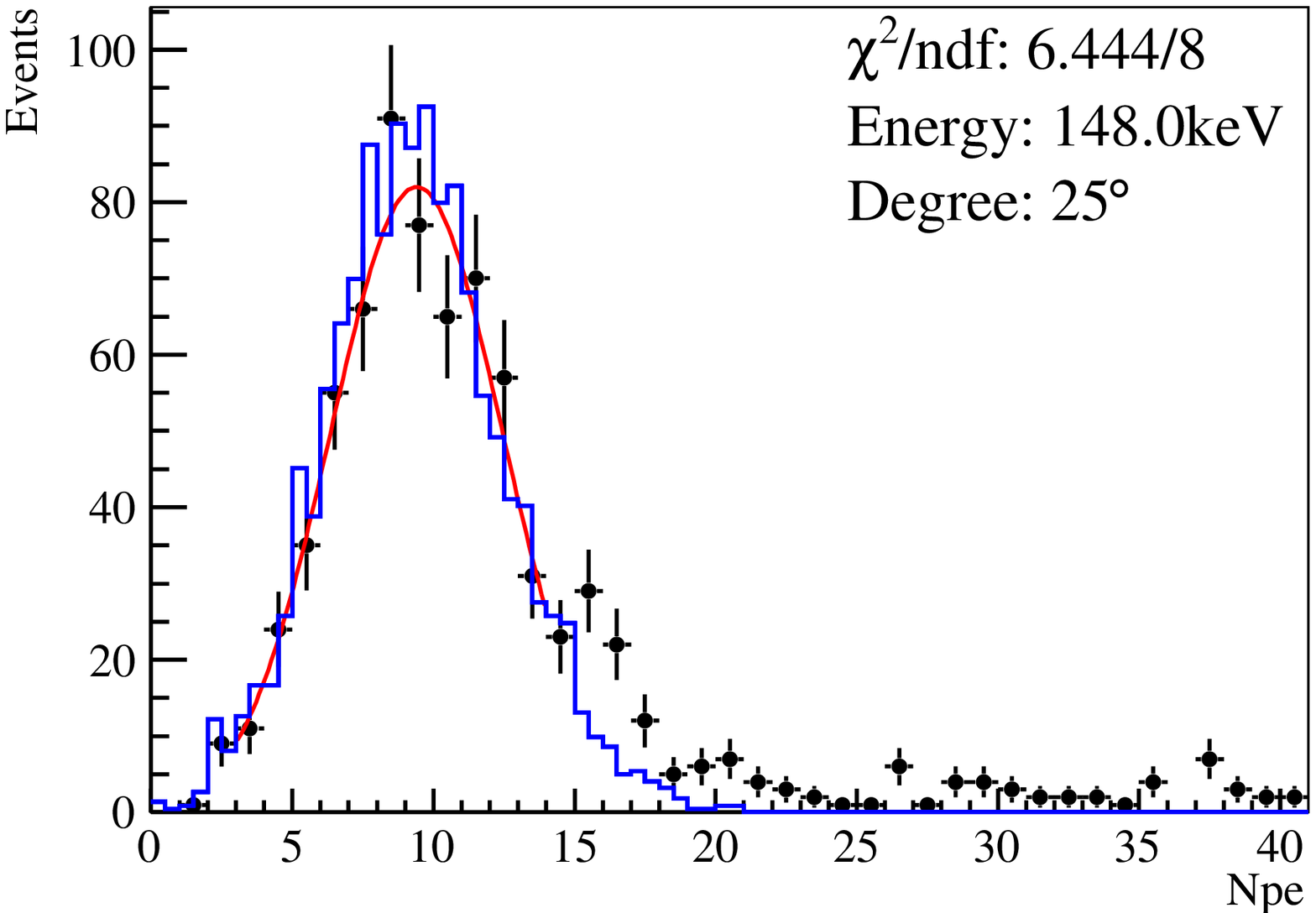}
  \includegraphics[width=3in]{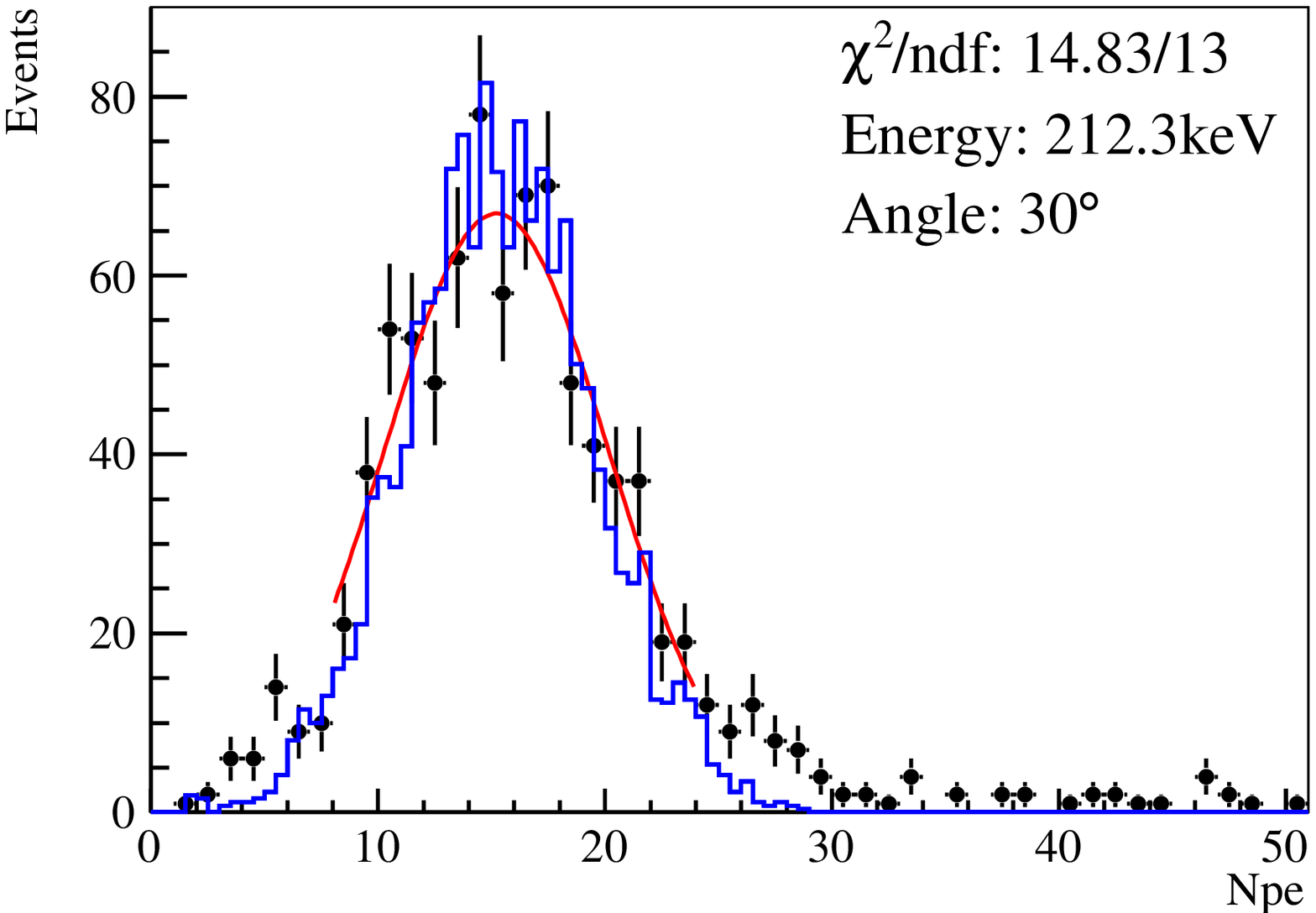}
  \includegraphics[width=3in]{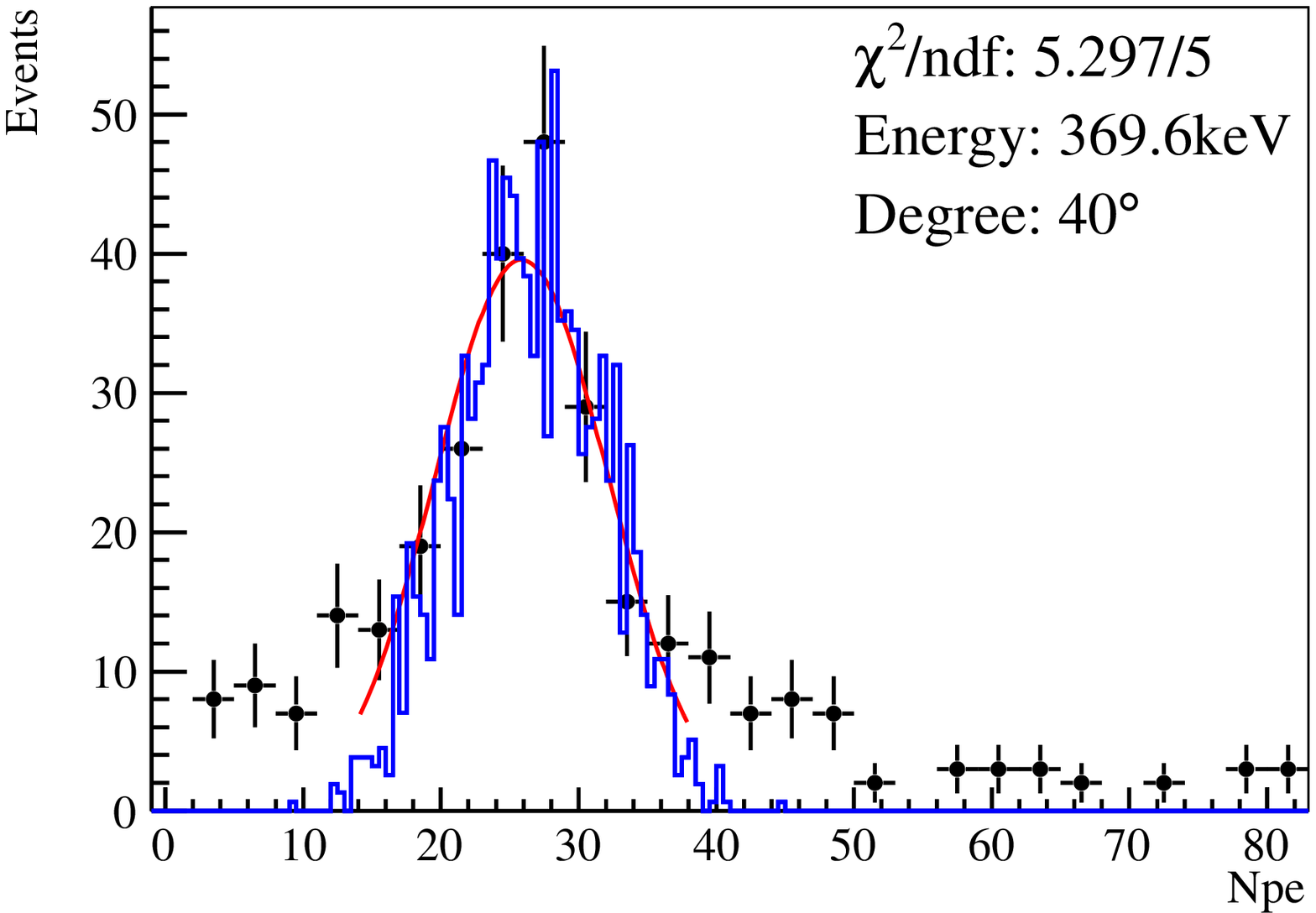}
    \caption{ Recoil energy spectra of BaF$_{2}$ tagged by each ND and fitted with Gaussian function. The labeled energies are F recoil energies. Black dots are experimental data and blue lines are Toy Monte Carlo simulations. }
  \label{spectrum}
 \end{figure*}

After the selection, the elastic scattering neutron events are clearly selected(Fig.~\ref{TOF}, upper plot, the red line). The number of photoelectron(Npe) distributions of elastic scattering events in BaF$_{2}$ are shown in Fig.~\ref{spectrum}. A Toy Monte Carlo is constructed to simulate the elastic scattering neutrons, including the effects of beam energy smear, detectors' geometry, NDs' efficiency and crystal response. The elastic scattering cross sections between nucleus and neutron are obtained from the National Nuclear Data Center database~\cite{nndc}. The simulation results are also shown in Fig.~\ref{spectrum}, which are basically consistent with the data. Because of large mass of Ba and low light yield of BaF$_{2}$, recoils from Ba at the three scattering angles are not observed.

\section{Results}

\begin{figure*}[!t]
  \centering
  \includegraphics[width=3in]{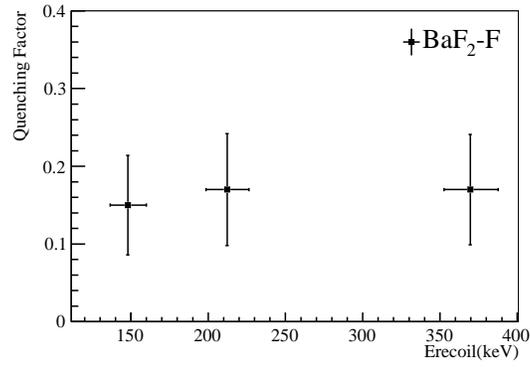}
     \caption{Quenching factors of F recoils in BaF$_2$. }
  \label{quenching}
 \end{figure*}

\begin{figure*}[!t]
  \centering
  \includegraphics[width=3in]{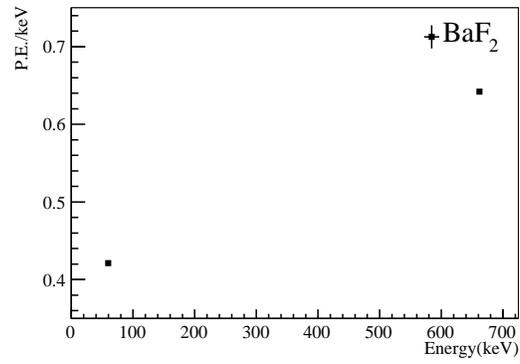}
     \caption{Light yield in different $\gamma$ ray energies of BaF$_2$. }
  \label{LY_difference}
 \end{figure*}

\begin{figure*}[!t]
  \centering
  \includegraphics[width=3in]{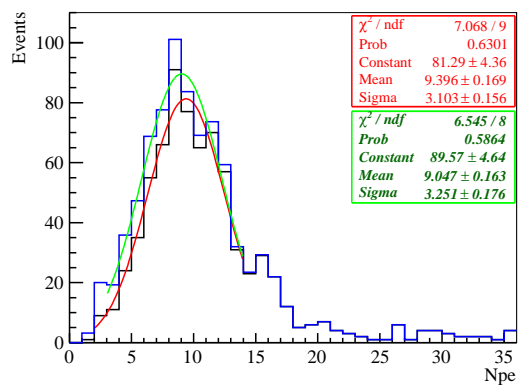}
     \caption{Comparison of the energy spectra with and without the efficiency correction for 25$^\circ$ ND selected events.}
  \label{BaF2_efficiency}
 \end{figure*}

The $\gamma$ sources, $^{241}Am$ and $^{137}Cs$, are used to calibrate BaF$_2$. Light yield of 0.42~p.e./keV is obtained. The quenching factor is defined as
\begin{eqnarray}
Q=\frac{E_{meas}}{E_{recoil}}
\label{quenching_def}
\end{eqnarray}
where $E_{meas}$ is the energy measured by the crystal, i.e., equivalent electron energy, and $E_{recoil}$ is the recoil energy corresponding to the scattering angle. The final quenching factors of the F recoils in BaF$_2$ is shown in Fig.~\ref{quenching}.  In Fig.~\ref{quenching} the horizontal uncertainties are dominated by the 1 degree
scattering angle uncertainties and the vertical one includes the contribution of the statistics, trigger efficiency and the systematics, which is dominated by crystal response non-linearity to electrons. In Eq.~\ref{quenching_def}, it is assumed that
the light yield is linear at different $\gamma$ energies, but the calibration data shows nonlinearity, indicating that the $\gamma$ also quenches in the crystal. The light yield differences for ${^{241}Am}$ and ${^{137}Cs}$ are taken as systematics, 41.6\% for BaF$_2$(Fig.~\ref{LY_difference}). Uncertainties from the trigger efficiency are also included in our measurements. The maximum shift of peak mean value based on the calculation for the lowest energy spectrum(Fig.~\ref{BaF2_efficiency}) is used as the uncertainty and is considered in the final results. Uncertainties from the efficiency are smaller effects than the effects from positioning and response non-linearity even for the lowest energy spectrum.

\begin{figure*}[!t]
  \centering
  \includegraphics[width=3in]{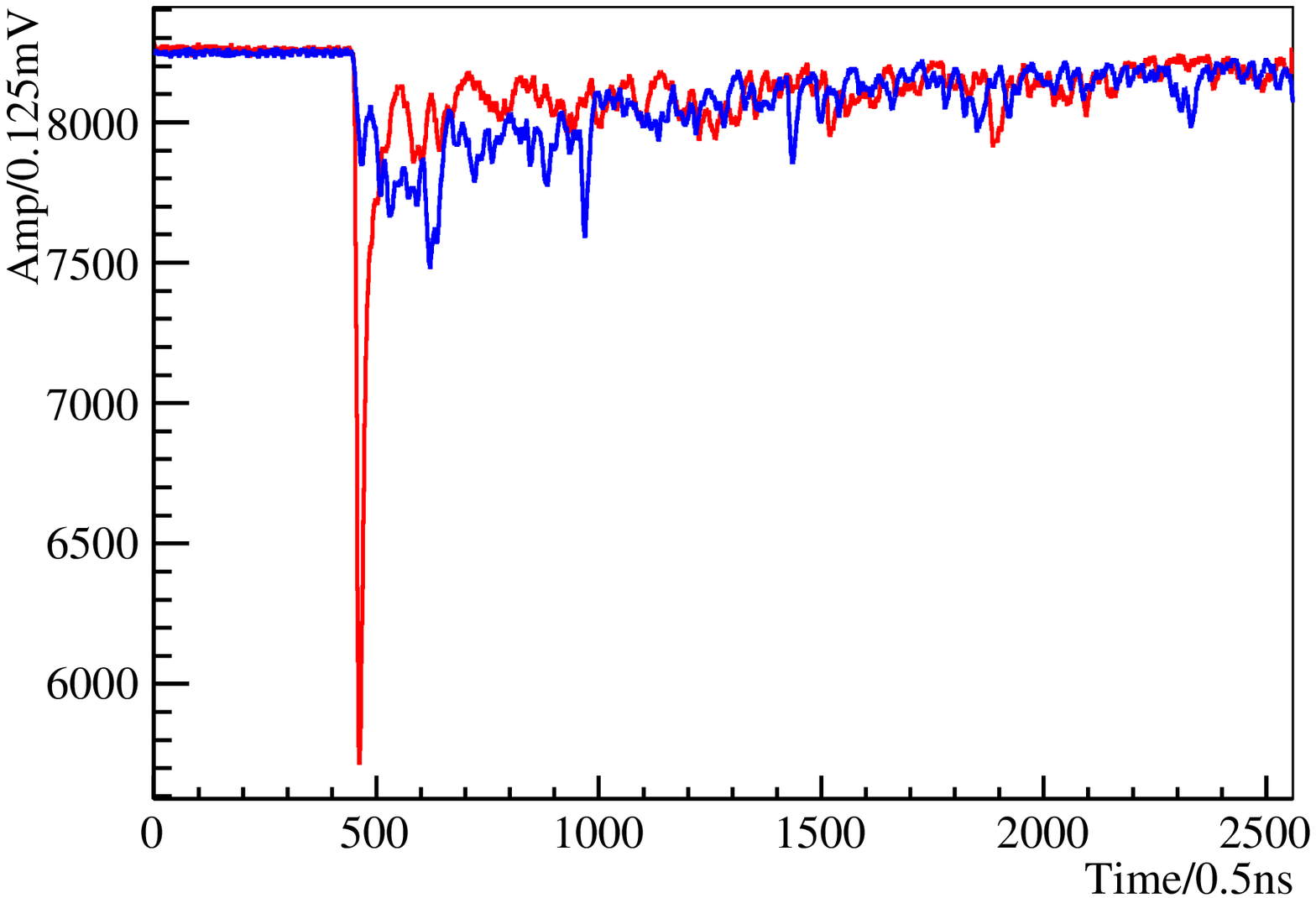}
  \includegraphics[width=3in]{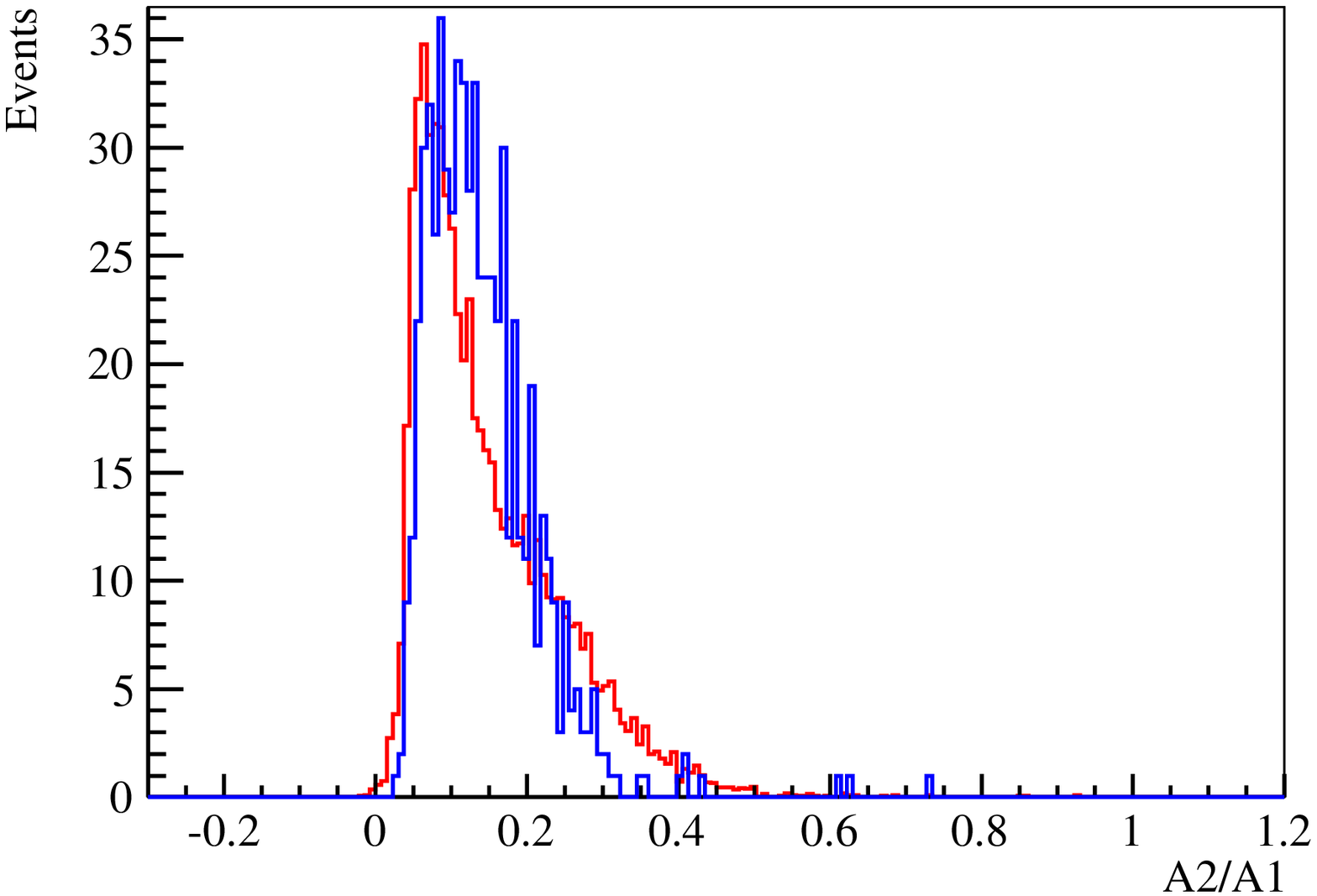}
  \includegraphics[width=3in]{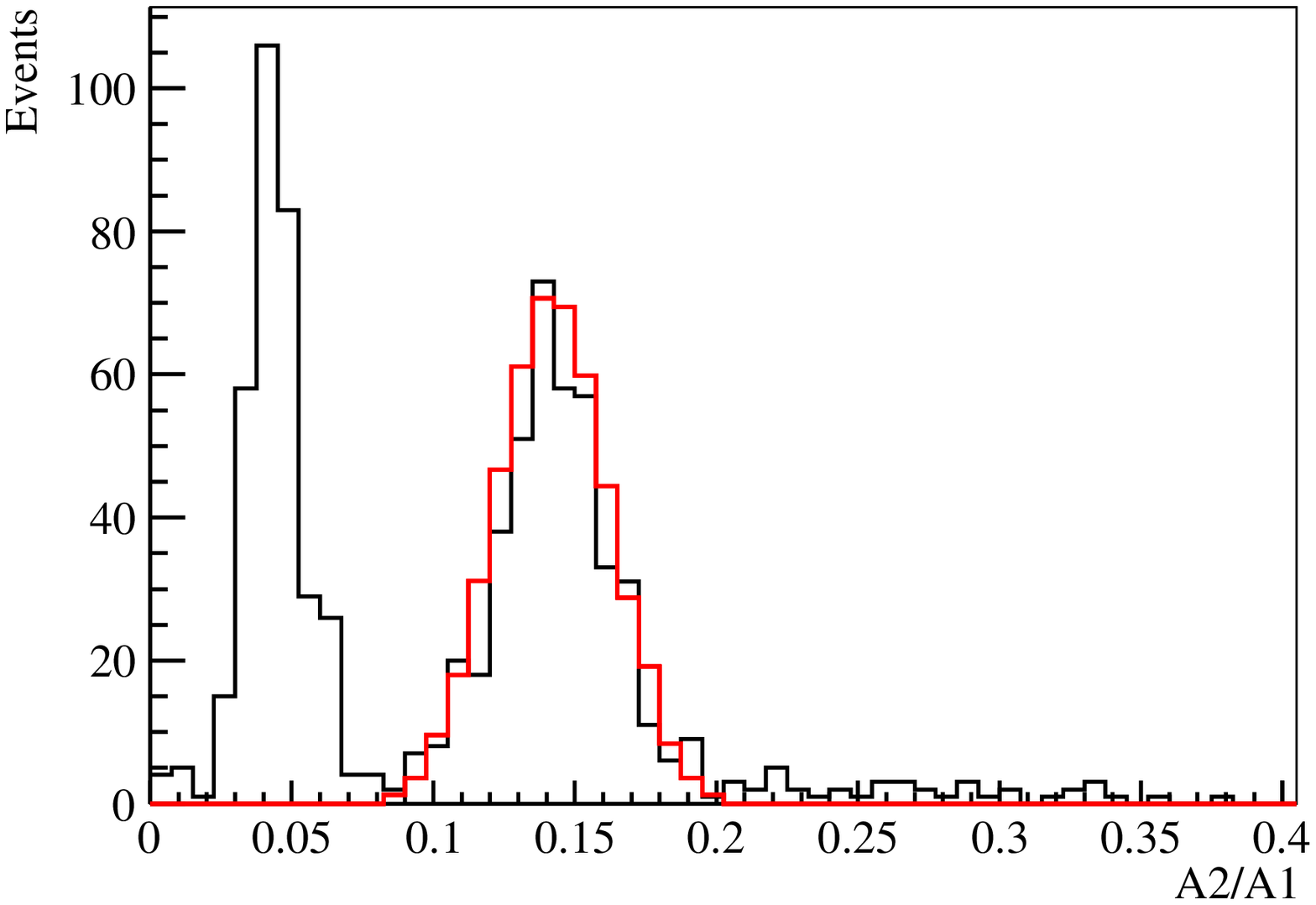}
    \caption{Discrimination between neutron and $\gamma$ of BaF$_{2}$. Upper: Example pulses of $\gamma$ events(red line) and $\alpha$ generated neutron inelastic scattering events(blue lines) having same p.e.s. Middle: A2/A1 distribution of elastic scattering neutron events triggered with ND1(blue line) and $\gamma$ events(red line). Lower: A2/A1 distribution of neutron inelastic scattering events(black line) and $\gamma$ events(red line) for 251 - 631~p.e. }
  \label{pulses}
 \end{figure*}

\begin{figure*}[!t]
  \centering
  \includegraphics[width=3in]{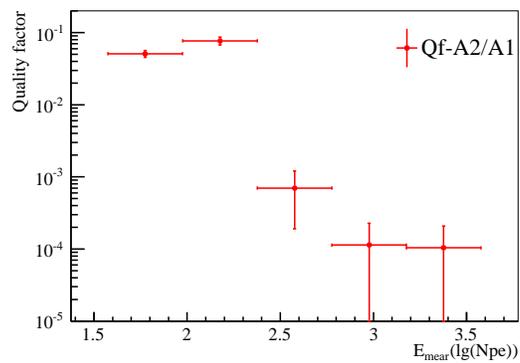}
    \caption{Quality factors of BaF$_2$ between neutron inelastic scattering with $\alpha$ generated events and $\gamma$ events. }
  \label{quality}
 \end{figure*}

Pulse shape discrimination technology is used to discriminate neutron and $\gamma$. Example pulses generated by $\gamma$ event and neutron inelastic scattering event in BaF$_{2}$ are presented in the Upper plot of Fig.~\ref{pulses}. The pulse of $\gamma$ event has obvious two components: the fast component and slow component. On the contrary, the pulse of neutron inelastic scattering event with $\alpha$ generated has only one slow component. The parameter $A_{2}$/$A_{1}$ is taken, where $A_{2}$ is the charge of the first 25~ns of the pulse and $A_{1}$ is the total charge of the pulse. The middle plot of Fig.~\ref{pulses} is the $A_{2}$/$A_{1}$ distribution for neutron elastic scattering events and $\gamma$ events, which indicates that it is hard to discriminate neutron elastic scattering events from $\gamma$ events at such low energies. The lower plot of Fig.~\ref{pulses} is the $A_{2}$/$A_{1}$ distribution for neutron inelastic scattering events and $\gamma$ events with 251 - 631~p.e. . The $A_{2}$/$A_{1}$ distribution of neutron inelastic scattering events have two peaks: the left one is from (n,$\alpha$) reaction on $^{19}F$ with cross section of 0.39 barn, and the right one is from (n,$\gamma$) reactions on $^{19}F$ with cross section of 0.24 barn. A quality factor is defined \cite{Gaitskell} as
\begin{eqnarray}
K {\equiv} \frac{{\beta}({1-\beta})}{({\alpha}-{\beta})^{2}}
\label{Eq:quenching}
\end{eqnarray}
where ${\alpha}$ is the fraction of signal events passing the event selection criteria and ${\beta}$ is the fraction of background events which pass the same criteria. For an ideal detector, ${\alpha}$ = 1 and  ${\beta}$ = 0. Therefore, a smaller quality factor means a better discrimination between signal and background events. The final results of quality factors between $\alpha$ generated neutron inelastic scattering events and $\gamma$ events are shown in Fig.~\ref{quality}. It indicates that neutron inelastic scattering events with $\alpha$ generated can be significantly discriminated from $\gamma$ events.

\section{Conclusions}

The nuclear recoils of BaF$_{2}$ crystal  is  studied  with  neutron beam. The energy spectra of nuclear recoils, quenching factors for elastic scattering neutrons and discrimination capability between neutron inelastic scattering events and ${\gamma}$ events are obtained for various recoil energies of the F content in BaF$_{2}$. The results indicate that BaF$_{2}$ is a good target material candidate for spin-dependent elastic scattering and inelastic scattering dark matter direct searching experiment.

\section{Acknowledgment}
This work is supported by the Ministry of Science and Technology of the People's Republic of China (No.~2010CB833003). We thank L.~Hou, H.T.~Chen and F.~Zhao of CIAE for their help during the experiment.





\bibliographystyle{model1-num-names}
\bibliography{<your-bib-database>}



\end{document}